\documentclass[twoside]{article}
\usepackage{qic,epsfig}

\renewcommand{\thefootnote}{\fnsymbol{footnote}}  
\usepackage{graphicx}
\usepackage{dcolumn}
\usepackage{bm}
\newcommand{\ket}[1]{\vert{#1}\rangle}

\begin{document}
\setlength{\textheight}{8.0truein}    

\runninghead{Near-Perfect Simultaneous Measurement of a Qubit Register}
            {M. Acton, K.-A. Brickman, P.C. Haljan, P.J. Lee, L. Deslauriers, C. Monroe}

\normalsize\textlineskip
\thispagestyle{empty}
\setcounter{page}{1}


\vspace*{0.88truein}

\alphfootnote

\fpage{1}

\centerline{\bf
NEAR-PERFECT SIMULTANEOUS MEASUREMENT OF A QUBIT REGISTER}
\vspace*{0.37truein}
\centerline{\footnotesize
M. Acton, K.-A. Brickman, P.C. Haljan\footnote{Present address: Department of Physics, Simon Fraser University, Burnaby, Canada.} , P.J. Lee\footnote{Present address: Laser Cooling and Trapping Group, National Institute of Standards and Technology, Gaithersburg, Maryland.} , L. Deslauriers\footnote{Present address: Department of Physics, Stanford University, Palo Alto, California.} , and C. Monroe}
\vspace*{0.015truein}
\centerline{\footnotesize\it FOCUS Center and Department of Physics, University
of Michigan}
\baselineskip=10pt
\centerline{\footnotesize\it 450 Church Street, Ann Arbor, MI 48109, USA}
\vspace*{0.225truein}
\publisher{November 29, 2005}{February 22, 2006}

\vspace*{0.21truein}

\abstracts{Simultaneous measurement of multiple qubits stored in hyperfine levels of trapped $^{111}$Cd$^+$ ions is realized with an intensified charge-coupled device (CCD) imager. A general theory of fluorescence detection for hyperfine qubits is presented and applied to experimental data. The use of an imager for photon detection allows for multiple qubit state measurement with detection fidelities of greater than $98\%$. Improvements in readout speed and fidelity are discussed in the context of scalable quantum computation architectures.}{}{}

\vspace*{10pt}

\keywords{simultaneous multiple ion qubit measurement, intensified CCD, state-dependent fluorescence theory}
\vspace*{3pt}

\vspace*{1pt}\textlineskip    


\setcounter{footnote}{0}
\renewcommand{\thefootnote}{\alph{footnote}}

\section{Introduction}
\noindent
Trapped atomic ions represent a promising method for implementing universal quantum computation. Trapped ions have already met most of the DiVincenzo requirements for quantum computation \cite{DiVincenzo00} and recent interest has focused on improvements in entangling gate fidelity \cite{Leibfried03,Schmidt03,Haljan05b,Lucas05}, producing large entangled states \cite{Liebfried05,Haffner05}, and scaling traps to larger numbers of ions \cite{Cirac00,Kielpinski02,Rowe02,Hensinger06}.

In this paper we discuss the important requirement of a multi-qubit measurement capability. In ions, state detection is accomplished by applying polarized laser light resonant with a cycling transition for one of the qubit states and off-resonant for the other state. The two states are then distinguishable as ``bright'' and ``dark'' via this state-dependent fluorescence \cite{Nagourney86,Sauter86,Bergquist86,Blatt88}. Typical schemes collect this fluorescence using fast lenses and detect photons using a standard photon-counting device such as a photo-multiplier tube (PMT) or an avalanche photo-diode (APD). The relatively high detection efficiency of PMTs or APDs aids detection, but for detecting more than one ion their lack of spatial resolution means that certain qubit states are indistinguishable, e.g.~one bright ion out of two does not determine a particular ion's state. Distinguishable individual qubit state detection is particularly crucial for tomographic density matrix reconstruction \cite{Roos04,Haljan05b}, quantum algorithms \cite{Chiaverini05,Brickman05}, quantum error correction \cite{Chiaverini04}, and cluster state quantum computation \cite{Raussendorf01,Duan05}.  Separating the ions with shuttling \cite{Kielpinski02,Rowe02,Hensinger06,Chiaverini05,Chiaverini04} or tightly focussing the detection beam \cite{Roos04} can distinguish the qubits, but the additional time necessary for detection, possible decoherence associated with shuttling, and technical difficulties make these schemes less desirable for large numbers of ions.

In this paper we discuss the use of an intensified charge-coupled device (CCD) as a photon-counting imager for simultaneously detecting multiple qubit states with high efficiency. We theoretically model the detection fidelity of qubits stored in $S_{1/2}$ hyperfine states of alkali-like ions, where one of the qubit states has a closed transition to the excited electronic $P$ state manifold (applicable to odd isotopes of Be$^+$, Mg$^+$, Zn$^+$, Cd$^+$, Hg$^+$, and Yb$^+$).  We present data for the detection of several $^{111}$Cd$^+$ ions using a CCD imager, and discuss technical features and limitations of current CCD technology. We finish with a discussion of future improvements and prospects for integration with scalable quantum computation architectures.

\section{Detection Theory} \label{detecttheory}
\noindent
\subsection{Basic detection method}
\noindent
There are two classes of alkali-like atomic ions that are amenable to high-fidelity $S_{1/2}$ hyperfine-state qubit detection. Ions that do not have a closed transition to the excited electronic $P$ state require shelving of one of the hyperfine qubit states to a low-lying metastable electronic $D$ state (odd isotopes of Ca$^+$, Sr$^+$, and Ba$^+$).  The detection efficiencies in this case can be very high; typically this method requires a narrowband laser source for high-fidelity shelving \cite{Roos99,Barton00} although recent work using rapid adiabatic passage may relax this laser requirement \cite{Wunderlich05}. Alternatively, one can obtain moderate detection efficiency by using coherent population trapping to optically shelve a particular spin state \cite{McDonnell04}. Ions that possess a closed transition to the excited electronic $P$ state (odd isotopes of Be$^+$, Mg$^+$, Zn$^+$, Cd$^+$, Hg$^+$, and Yb$^+$) can be detected directly, and will be the focus of this paper. Throughout this paper we assume that the Zeeman splitting is small compared to the hyperfine splittings.

\begin{figure} [htbp]
\centerline{\epsfig{file=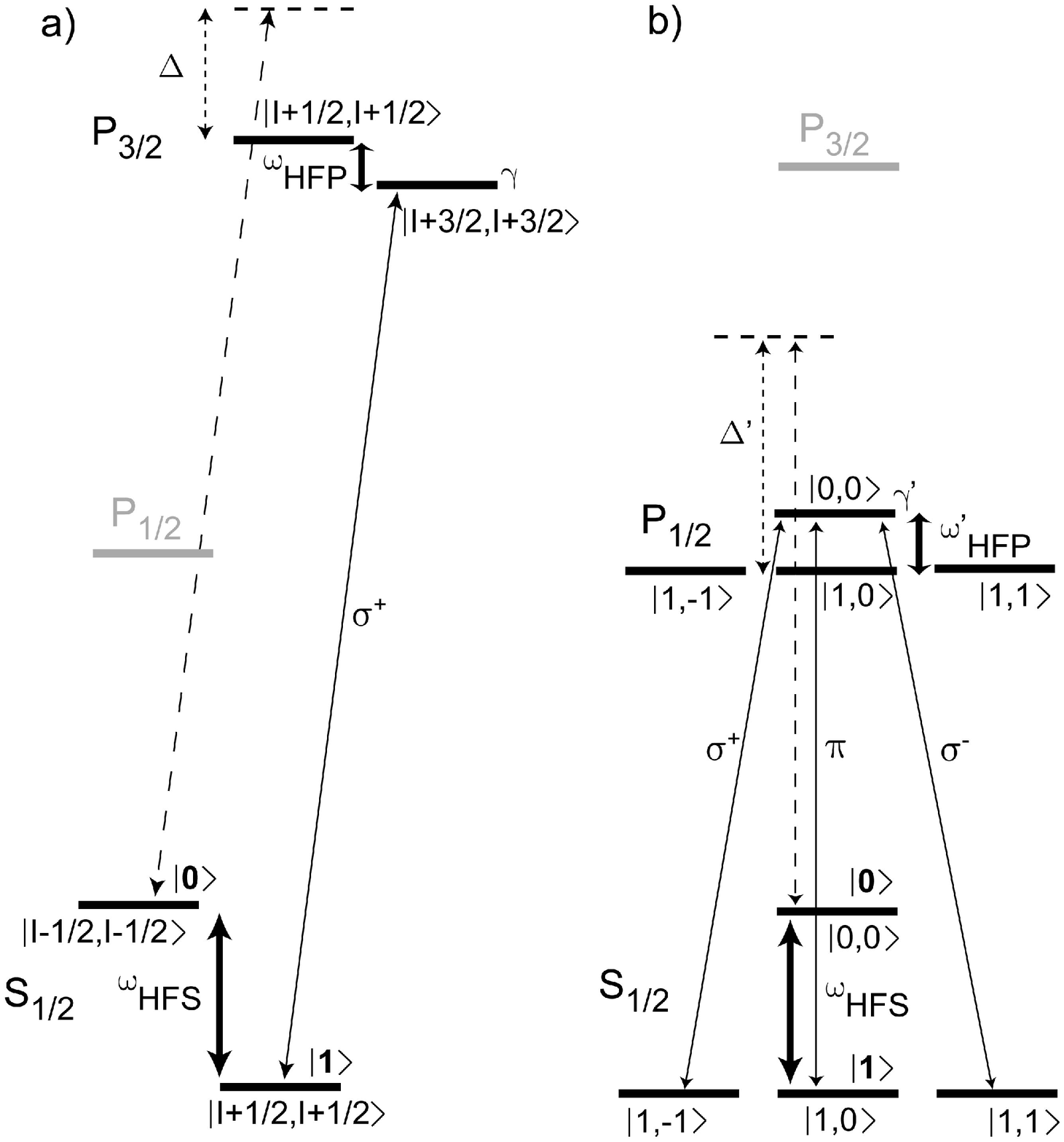, width=4.0in}}
\vspace*{13pt}
\fcaption{\label{genenergylevels}Relevant energy levels used in fluorescence detection of qubits stored in the hyperfine levels of the $S_{1/2}$ ground state of atoms with a single valence electron, with no relevant low-lying excited states below the excited $P$ manifold.  Energy levels are labeled by the $\ket{F,m_F}$ quantum numbers of total angular momentum and the energy splittings are not to scale. a) Detection through $P_{3/2}$ level with qubit stored in the $S_{1/2} \ket{I+1/2,I+1/2} \equiv \ket{{\bf 1}}$ and $S_{1/2} \ket{I-1/2,I-1/2} \equiv \ket{{\bf 0}}$ ``stretch'' hyperfine states, for any nonzero nuclear spin $I$.  By applying $\sigma^+$-polarized laser radiation resonant with the $\ket{{\bf 1}} \to P_{3/2} \ket{I+3/2,I+3/2}$ cycling transition, qubit state $\ket{{\bf 1}}$ results in strong fluorescence, while qubit state $\ket{{\bf 0}}$ is nearly dark owing to a detuning of $\Delta=\omega_{\mathrm{HFS}}-\omega_{\mathrm{HFP}} \gg \gamma$ to the nearest resonance, where $\omega_{\mathrm{HFS}}$ and $\omega_{\mathrm{HFP}}$ are the hyperfine splittings of the $S_{1/2}$ and $P_{3/2}$ states and $\gamma$ is the radiative linewidth of the $P_{3/2}$ state. b) Detection through the $P_{1/2}$ level with qubit stored in the $S_{1/2} \ket{1,0} \equiv \ket{{\bf 1}}$ and $S_{1/2} \ket{0,0} \equiv \ket{{\bf 0}}$ ``clock'' hyperfine states for the special case of nuclear spin $I=1/2$.  Here, applying all polarizations of laser radiation resonant with the $\ket{{\bf 1}} \to P_{1/2} \ket{0,0}$ transition results in strong fluorescence, while qubit state $\ket{{\bf 0}}$ is nearly dark owing to a detuning of $\Delta'=\omega_{\mathrm{HFS}}+\omega_{\mathrm{HFP}}' \gg \gamma'$ to the nearest resonance, where $\omega_{\mathrm{HFS}}$ and $\omega_{\mathrm{HFP}}'$ are the hyperfine splittings of the $S_{1/2}$ and $P_{1/2}$ states and $\gamma'$ is the radiative linewidth of the $P_{1/2}$ state.}
\end{figure}

There are two basic schemes for this direct state detection as outlined in figure~\ref{genenergylevels}. For both methods, the qubit is stored in the hyperfine levels of the $S_{1/2}$ manifold with hyperfine splitting $\omega_{\mathrm{HFS}}$. Discussing the general case first (fig.~\ref{genenergylevels}a), if we write the states in the $\ket{F,m_F}$ basis with $I$ the nuclear spin, the $S_{1/2} \ket{I+1/2,I+1/2}\equiv \ket{{\bf 1}}$ state exhibits a closed ``cycling'' transition to the $P_{3/2} \ket{I+3/2,I+3/2}$ state when resonant $\sigma^+$-polarized laser light is applied\footnote{Equivalently one could use $\sigma^-$-polarized radiation with appropriate qubit and excited states.}. If the qubit is in the $\ket{{\bf 1}}$ state then the resonant laser light induces a large amount of fluorescence. When a portion of these photons are collected and counted on a photon-counting device, a histogram of their distribution follows a Poissonian distribution with a mean number of collected photons that is determined by the laser intensity and application time, the upper-state radiative linewidth $\gamma$, and the photon collection efficiency of the detection system. In contrast, when the qubit is in the $S_{1/2} \ket{I-1/2,I-1/2}\equiv \ket{{\bf 0}}$ state the laser radiation is no longer resonant with the transition to any excited state. The nearest allowed transition is to $P_{3/2} \ket{I+1/2,I+1/2}$ which is detuned by $\Delta=\omega_{\mathrm{HFS}}-\omega_{\mathrm{HFP}}$, where $\omega_{\mathrm{HFP}}$ is the hyperfine splitting of the $P_{3/2}$ states, so an ion in the $\ket{{\bf 0}}$ state scatters virtually no photons. Thus, we can determine the qubit's state with high fidelity by applying $\sigma^+$-polarized laser radiation resonant with the cycling transition and counting the number of photons that arrive at the detector.

For ions with isotopes that have nuclear spin $I=1/2$ (Cd$^+$, Hg$^+$, and Yb$^+$), there is another possible state-dependent fluorescence detection mechanism by coupling to the $P_{1/2}$ manifold (fig.~\ref{genenergylevels}b). If we apply all polarizations of laser light ($\sigma^+$, $\pi$, and $\sigma^-$) resonant with the $S_{1/2} \ket{F=1}\to P_{1/2} \ket{F=0}$ transition then the only allowed decay from the excited state is back to the $F=1$ levels of $S_{1/2}$ \cite{Bergquist02}, forming a closed cycling transition. If the ion begins in the state $\ket{1,0}\equiv \ket{{\bf 1}}$, the ion will fluoresce many photons under this laser stimulation and we can collect these photons as above. Conversely, the state $\ket{0,0}\equiv \ket{{\bf 0}}$ will scatter virtually no photons under this laser light because it is off-resonant from its only allowed transition to the $P_{1/2} \ket{F=1}$ levels by $\Delta'=\omega_{\mathrm{HFS}}+\omega_{\mathrm{HFP}}'$, where $\omega_{\mathrm{HFP}}'$ is the hyperfine splitting of the $P_{1/2}$ levels. Note that to avoid an optically-pumped dark state formed by a coherent superposition of $S_{1/2} \ket{1,-1}$, $\ket{1,0}$, and $\ket{1,1}$ it is necessary to modulate the laser polarization or use a magnetic field to induce a well-chosen Zeeman splitting \cite{Berkeland02}.

In the following sections we present a general theory of this state-dependent fluorescence by determining the off-resonant coupling between the qubit states. We quantify these detection errors in order to calculate the fidelity of qubit state detection for various photon detection efficiencies.

\subsection{Statistics: dark $\to$ bright leakage}
\noindent
For both the general and the $I=1/2$ specific detection methods, qubits in the dark state can leak onto the bright transition by off-resonantly coupling to the wrong hyperfine excited level during detection. Rate equations describing this off-resonant pumping yield an exponential probability distribution of remaining in the dark state as a function of time.  Once in the bright state, the collected photons from the closed transition obey Poissonian statistics.  Therefore, for a qubit initially in the dark state, we expect the distribution of emitted photons to be a convolution of Poissonian and exponential distributions \cite{Kingthesis,Roosthesis}, as we now derive.

The probability of leaving the dark state at a time $t$ is given by:
\begin{equation} \label{darktimedecay}
f(t)\textrm{d}t=\frac{1}{\tau_{L1}}e^{-\frac{t}{\tau_{L1}}}\textrm{d}t
\end{equation}
where $\tau_{L1}$ is the average leak time of the dark state onto the closed transition. Also, the average number of collected photons for a qubit that starts dark but is pumped to a bright state at time $t$ is:
\begin{equation} \label{darkmean}
\lambda(t)=(1-\frac{t}{\tau_D})\lambda_0
\end{equation}
where $\tau_D$ is the detection time and $\lambda_0$ is the mean number of {\it counted} photons when starting in the bright state. 

We want to transform from a probability distribution $f(t)\textrm{d}t$ to a probability distribution of Poissonian means $g(\lambda)\textrm{d}\lambda$ so we use eqn.~\ref{darkmean} to get $t(\lambda)$ and then substitute into eqn.~\ref{darktimedecay}. This yields the probability of the dark qubit state producing a Poissonian distribution of collected photons with mean $\lambda$:
\begin{equation} \label{darkmeanprob}
g(\lambda)\textrm{d}\lambda=\left\{ \begin{array}{ll} 
	\frac{\alpha_1}{\eta} e^{(\lambda-\lambda_0) \alpha_1/\eta} \textrm{d}\lambda & \lambda > 0\\
	e^{-\alpha_1 \lambda_0/\eta} & \lambda = 0
\end{array} \right.
\end{equation}
where $\eta=\eta_D \frac{\textrm{d}\Omega}{4\pi}T$ is the total photon collection efficiency determined by the detector efficiency ($\eta_D$), the solid angle of collection ($\frac{\textrm{d}\Omega}{4\pi}$), and the optical transmission from the ion to the detector ($T$); $\alpha_1 \equiv \frac{\tau_D \eta}{\tau_{L1} \lambda_0}$ is the leak probability per {\it emitted} photon; and the $\lambda=0$ discontinuity is necessary to account for the fraction that do not leave the dark state.

Therefore, the probability of detecting $n$ photons when starting in the dark state is the convolution of $g(\lambda)$ with the Poissonian distribution $P(n|\lambda)=\frac{e^{-\lambda}\lambda^n}{n!}$:
\begin{equation} \label{darkprobinit}
p_{\mathrm{dark}}(n) = \delta_n e^{-\alpha_1 \lambda_0/\eta} + \int_{\epsilon}^{\lambda_0} \frac{e^{-\lambda} \lambda^{n}}{n!} \frac{\alpha_1}{\eta} e^{(\lambda-\lambda_0) \alpha_1/\eta} \textrm{d}\lambda
\end{equation}
with $\delta_n$ the Kronecker delta function and $\epsilon \to 0$. Re-writing in terms of the incomplete Gamma function we obtain:
\begin{equation} \label{darkprobfinal}
p_{\mathrm{dark}}(n) = e^{-\alpha_1 \lambda_0/\eta} \left[ \delta_n + \frac{\alpha_1/\eta}{(1-\alpha_1/\eta)^{n+1}}\mathcal{P}(n+1,(1-\alpha_1/\eta)\lambda_0)\right]
\end{equation}
where $\mathcal{P}(a,x)\equiv \frac{1}{(a-1)!}\int_{0}^x e^{-y}y^{a-1}\textrm{d}y$ is the standard definition of the incomplete Gamma function normalized such that $\mathcal{P}(a,\infty)=1$.

Figure~\ref{genhistograms}a shows this probability distribution for various values of the leakage parameter, $\alpha_1$. Later, we will calculate physical values of $\alpha_1$ and $\lambda_0$ from the atomic parameters.

\subsection{Statistics: bright $\to$ dark leakage}
\begin{figure} [htbp]
\centerline{\epsfig{file=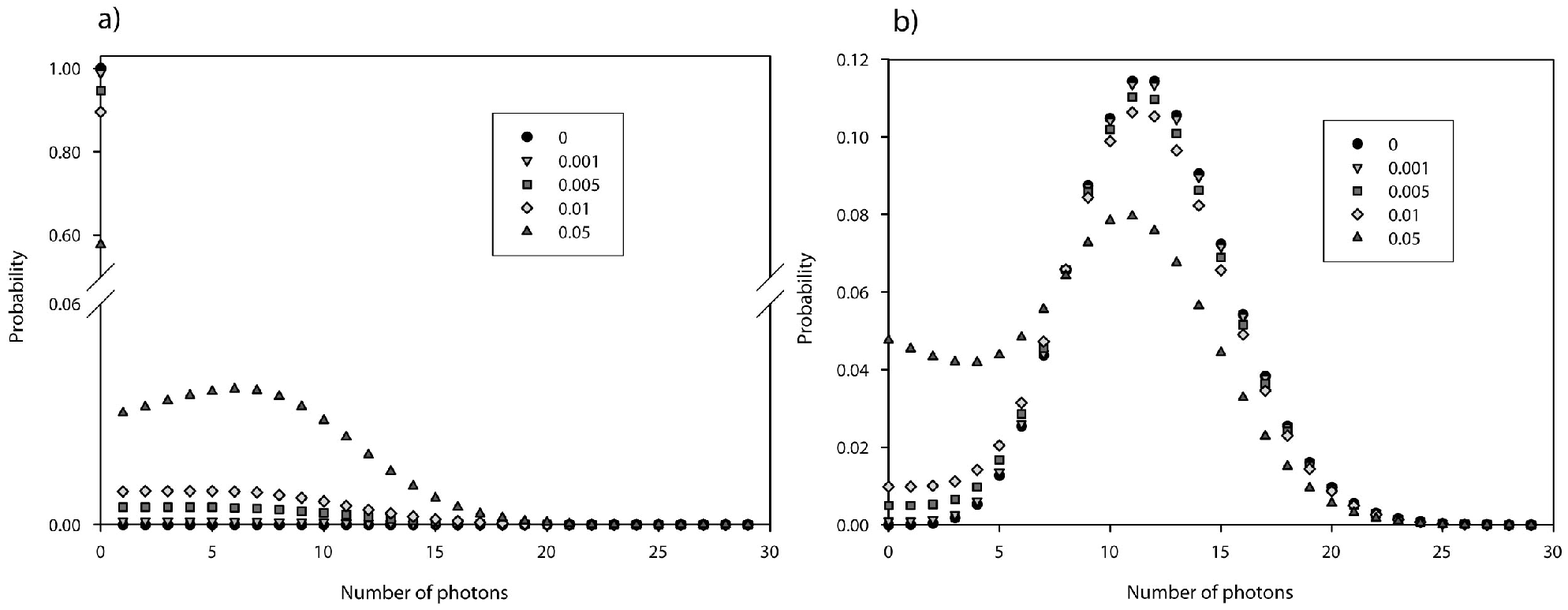, width=\columnwidth}}
\vspace*{13pt}
\fcaption{\label{genhistograms}Theoretical photon count histograms for ions that start in: a) the nominally ``dark'' $\ket{{\bf 0}}$ state; and b) the nominally ``bright'' $\ket{\bf{1}}$ state. Both plots use $\lambda_0=12$ with varying values of: a) $\alpha_1/\eta$ (the leakage probability from the dark state to the bright state per {\it detected} photon) and b) $\alpha_2/\eta$ (the leakage probability from the bright state to the dark state per {\it detected} photon).}
\end{figure}

\noindent
Similarly, the bright state can be optically pumped into the dark state by off-resonant coupling to the wrong excited hyperfine level. When using the $P_{3/2}$ manifold for detection this coupling can only occur for imperfect laser polarizations, while for the specialized $I=1/2$, $P_{1/2}$ case this off-resonant coupling is always present (since all laser polarizations are applied). As before, the overall photon probability distribution will be a convolution of Poissonian and exponential distributions, but now reversed: after some time scattering photons on the closed transition the qubit is pumped into the dark state and emits no more photons (neglecting the 2nd order effect of then re-pumping from the dark state back to the bright state). If we define the leak probability per emitted photon $\alpha_2 \equiv \frac{\tau_D \eta}{\tau_{L2} \lambda_0}$ where $\tau_{L2}$ is the average leak time from the cycling transition into the dark state, then with similar statistics as before, the probability of detecting $n$ photons when starting in the bright state is:
\begin{equation} \label{brightprobfinal}
p_{\mathrm{bright}}(n) = \frac{e^{-(1+\alpha_2/\eta)\lambda_0} \lambda_0^n}{n!} + \frac{\alpha_2/\eta}{(1+\alpha_2/\eta)^{n+1}}\mathcal{P}(n+1,(1+\alpha_2/\eta)\lambda_0)
\end{equation}
where the first term is the Poissonian distribution from never leaving the closed transition and the second term is the smearing of the distribution from pumping to the dark state (note that $\alpha_1 \neq \alpha_2$). This distribution is shown in figure~\ref{genhistograms}b for varying values of $\alpha_2$.

One can further separate the dark and bright distributions by using ancilla qubits and entangling gates before measurement \cite{Schaetz05}. For ions with large leak probabilities $\alpha_1$ or $\alpha_2$ this technique may be useful because it can increase qubit detection fidelity to as high as the entangling gate fidelity.

\subsection{Calculating the atomic parameters}
\noindent
\subsubsection{General case using $P_{3/2}$} \label{gendetectparams}
\noindent
If we assume a detuning $\delta$ of the detection beam from the cycling transition resonance then for ideal $\sigma^+$-polarized laser radiation the mean number of photons {\it detected} from a qubit in $\ket{{\bf 1}}$ is:
\begin{equation} \label{cyclingmean}
\lambda_0 = \tau_D \eta \frac{s\frac{\gamma}{2}}{1+s+\left(\frac{2\delta}{\gamma}\right)^2}
\end{equation}
where $\tau_D$ is the detection time; $\eta$ is the {\it total} end-to-end efficiency of photon detection; $\gamma$ is the radiative linewidth of the $P_{3/2}$ state; and $s=I/I_{\mathrm{sat}}$ is the laser-ion saturation parameter.

Next we can calculate $\alpha_1$, the probability per emitted photon of leakage from the dark state into the bright state. The dark state can off-resonantly couple to the $P_{3/2} \ket{I+1/2,I+1/2}$ level with probability that goes as: $\sim s \left(\frac{\gamma}{2\Delta_1}\right)^2$, where $\Delta_1=\omega_{\mathrm{HFS}}-\omega_{\mathrm{HFP}}$ is the $S_{1/2}$ hyperfine splitting minus the $P_{3/2}$ hyperfine splitting between $\ket{I+3/2,I+3/2}$ and $\ket{I+1/2,I+1/2}$. We also need to take into account the branching ratio, ${\bf M_1}$, to determine the transition rate between the dark state and the cycling transition. Thus, we can write the value of $\alpha_1$ as:
\begin{equation} \label{darkalpha}
\alpha_1 = {\bf M_1}(1+s+\left(\frac{2\delta}{\gamma}\right)^2) \left(\frac{\gamma}{2\Delta_1}\right)^2.
\end{equation}

We can calculate the braching ratio, ${\bf M_1}$, by considering off-resonant coupling between the dark state ($\ket{I-1/2,I-1/2}$) and the state $P_{3/2} \ket{I+1/2,I+1/2}$. This state can then decay into the bright manifold ground states $\ket{I+1/2,I \pm 1/2}$. The generalized formula can be written as:
\begin{eqnarray} \label{m1ratio}
{\bf M_1} &=& C(I-1/2,I+1/2;I-1/2,I+1/2) \sum_{i=I-1/2}^{I+1/2} C(I+1/2,I+1/2;I+1/2,i) \nonumber\\
&=&\frac{4I(3+2I)}{9(1+2I)^2}.
\end{eqnarray}
where we have used that $C(F,F';f,f')$ is the square of the Clebsch-Gordon coefficient between two states $F \to F'$ and $m_F=f \to f'$ given by \cite{Metcalf}:
{\setlength\arraycolsep{2pt}
\begin{eqnarray} \label{CGcoeff}
C(F,F';f,f') & = & \left[(2J+1)(2J'+1)(2F+1)(2F'+1) \right] \nonumber\\
& & \times \left[ \left\{ \begin{array}{ccc}
L' & J' & S \\
J & L & 1 \end{array} \right\}
\left\{ \begin{array}{ccc}
J' & F' & I \\
F & J & 1 \end{array} \right\}
\left( \begin{array}{ccc}
F & 1 & F' \\
f & q & -f' \end{array} \right) \right]^2
\end{eqnarray}}
where $\{ \}$ is the 6-J symbol, $( )$ is the 3-J symbol with polarization number $q$, and there is an implied normalization constant such that the cycling transition strength is 1.

Similarly, we can calculate $\alpha_2$, the probability per emitted photon of leakage from the bright state into the dark state. Note that there are two leakage paths out of the bright state: via coupling to the $P_{3/2} \ket{I+1/2,I+1/2}$ state (due to $\pi$-polarized laser light) or to the $P_{3/2} \ket{I+1/2,I-1/2}$ state (due to $\sigma^{-}$-polarized radiation). These two paths yield a leakage probability:
\begin{equation} \label{brightalpha}
\alpha_2 = (1+s+\left(\frac{2\delta}{\gamma}\right)^2) \left(\frac{\gamma}{2\Delta_2}\right)^2\frac{{\bf M_{2\pi}}P_{\pi}+{\bf M_{2-}}P_{-}}{1-(P_{\pi}+P_{-})}
\end{equation}
where $\Delta_2=\omega_{\mathrm{HFP}}$ is the hyperfine splitting of the $P_{3/2}$ levels, $P_{\pi}$ ($P_{-}$) is the fraction of $\pi$ ($\sigma^{-}$)-polarized laser power, and ${\bf M_{2\pi}}$ (${\bf M_{2-}}$) is the dipole branching ratio for $\pi$ ($\sigma^{-}$)-polarized light impurity. These branching ratios are given by:
\begin{eqnarray} \label{m2pratio}
{\bf M_{2\pi}} &=& C(I+1/2,I+1/2;I+1/2,I+1/2) \times C(I-1/2,I+1/2;I-1/2,I+1/2) \nonumber\\
&=& \frac{4I}{9+18I}
\end{eqnarray}
and
\begin{eqnarray} \label{m2sratio}
{\bf M_{2-}} &=& C(I+1/2,I+1/2;I+1/2,I-1/2) \times C(I-1/2,I+1/2;I-1/2,I-1/2) \nonumber\\
&=& \frac{16I}{9(1+2I)^3}.
\end{eqnarray}

\subsubsection{$I=1/2$ case using $P_{1/2}$}
\noindent
For the special case of detecting the qubit state via the $P_{1/2}$ manifold, we need to make some small adjustments to our calculations. Assuming that the laser power is split roughly equally between the 3 polarizations, the mean number of collected, scattered photons when starting in one of the bright, $F=1$ states is:
\begin{equation}
\lambda_0' = \tau_D \eta \frac{s \frac{\gamma'}{2}}{1+s+\left(\frac{2\delta}{\gamma'}\right)^2}
\end{equation}
where $\gamma'$ is the radiative linewidth of the $P_{1/2}$ state.

For the dark state leakage per emitted photon, $\alpha_1'$, we simply have a new relevant detuning $\Delta_1'$ and branching ratio ${\bf M_1'}$:
\begin{equation}
\alpha_1' = {\bf M_1'} (1+s+\left(\frac{2\delta}{\gamma'}\right)^2) \left(\frac{\gamma'}{2\Delta_1'}\right)^2.
\end{equation}
where the relevant detuning is now $\Delta_1'=\omega_{\mathrm{HFS}}+\omega_{\mathrm{HFP}}'$, namely the $S_{1/2}$ state hyperfine splitting plus the $P_{1/2}$ state hyperfine splitting. For the $I=1/2$ ions, all allowed dipole transitions have relative strength of $1/3$, so the dark state leakage branching ratio is: ${\bf M_1'}=1/3\times(1/3+1/3)=2/9$.

Also, the bright state can leak into the dark state via off-resonant coupling to one of the $P_{1/2}$, $F=1$ states with probability per emitted photon of:
\begin{equation}
\alpha_2' = {\bf M_2'} (1+s+\left(\frac{2\delta}{\gamma'}\right)^2) \left(\frac{\gamma'}{2\Delta_2'}\right)^2
\end{equation}
where $\Delta_2'=\omega_{\mathrm{HFP}}'$, the hyperfine splitting of the $P_{1/2}$ state, and ${\bf M_2'}=(1/3+1/3)\times 1/3=2/9$.

\subsection{Example calculation using $^{111}$Cd$^+$}
\noindent
We now compare the above model to experimental data for $^{111}$Cd$^+$ qubits using a PMT for photon detection. For this calculation and experiment we use the general detection technique that couples to the $P_{3/2}$ levels.

\begin{figure} [htbp]
\centerline{\epsfig{file=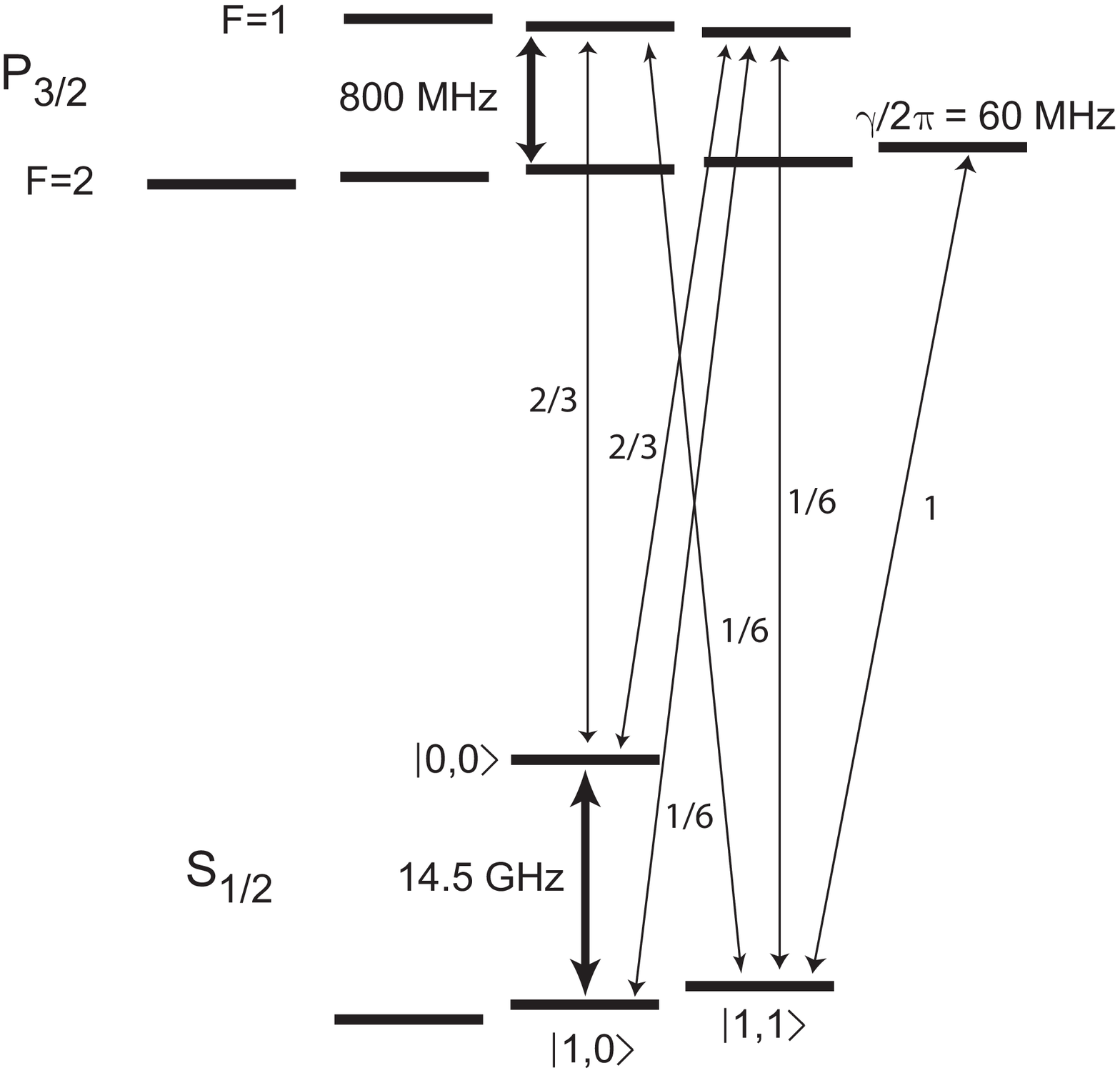, width=3.0in}}
\vspace*{13pt}
\fcaption{\label{dipolestrengths}Relevant energy level spacings and relative electric dipole strengths for selected transitions in $^{111}$Cd$^+$ \cite{Tanaka96,Moehring06}. Note that the Zeeman splitting ($< 10$~MHz) is small compared to the hyperfine splittings.}
\end{figure}

\begin{figure} [htbp]
\centerline{\epsfig{file=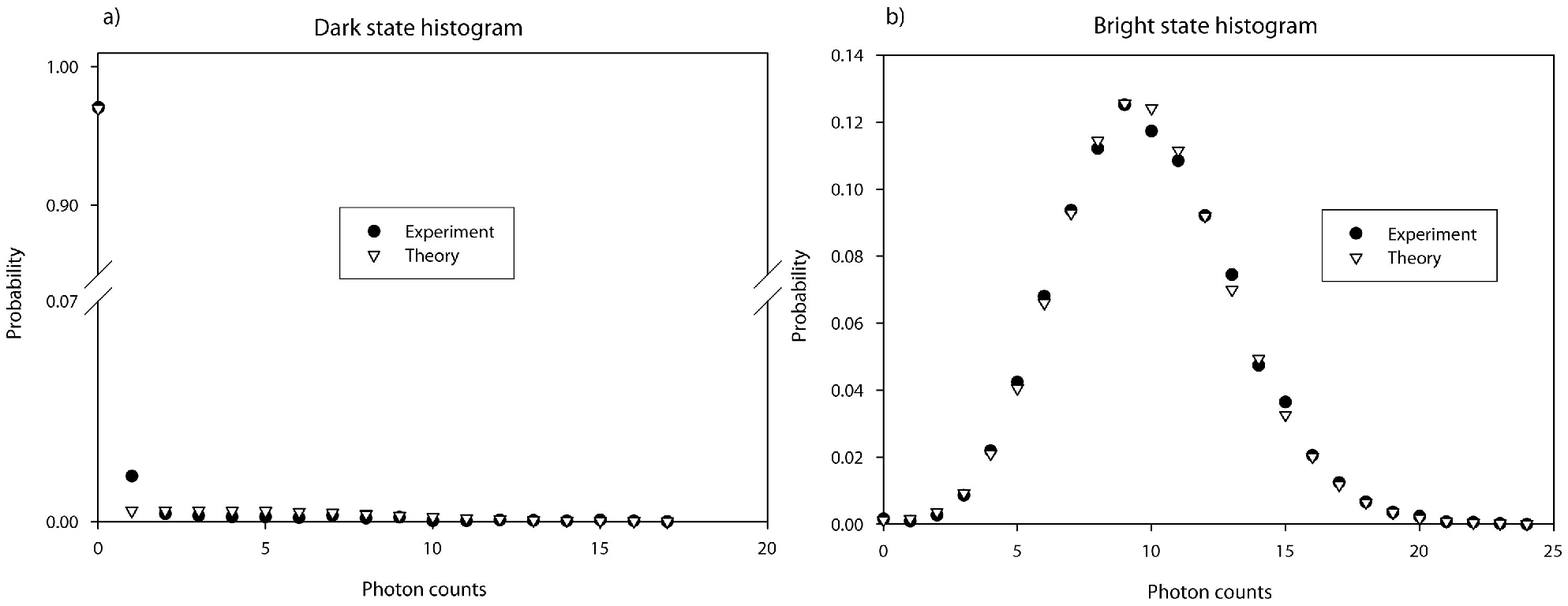, width=\columnwidth}}
\vspace*{13pt}
\fcaption{\label{detectfit}Detection histograms using a PMT and the $P_{3/2}$ detection scheme for a single $^{111}$Cd$^+$ ion prepared in: a) $\ket{0,0}\equiv \ket{{\bf 0}}$ dark state; and b) $\ket{1,0}\equiv \ket{{\bf 1}}$ bright state. For the data shown, each state is prepared and then measured 20,000 times. Fit is to theory from text (eqn.~\ref{darkprobfinal} and eqn.~\ref{brightprobfinal}, respectively) with parameters: $\tau_D=150\mu$s, $\eta=1.4\times10^{-3}$, $P_{\mathrm{impure}}=1.5\times10^{-3}$, and $s=0.25$. Note that the $n=1$ bin in a) includes background laser light scatter that the model does not include.}
\end{figure}

In the experiment, individual $^{111}$Cd$^+$ ions ($I=1/2$) are confined in a linear RF Paul trap with controllable axial frequency $\omega_z/2\pi = 0.5-2.5 \textrm{~MHz}$ \cite{Deslauriers04}. The qubit is stored in the first-order magnetic field-insensitive ``clock'' states: $\ket{0,0}\equiv\ket{{\bf 0}}$ and $\ket{1,0}\equiv\ket{{\bf 1}}$. In $^{111}$Cd$^+$, for $\sigma^+$-polarized laser light the large hyperfine splitting of the $P_{3/2}$ states ensures that population in the $S_{1/2} \ket{1,0}$ state is optically pumped to the $S_{1/2} \ket{1,1}$ with near unit probability so the $P_{3/2}$ detection theory given in sec.~\ref{gendetectparams} will apply well to this case (corrections are given in sec.~\ref{theoryfidelity}). For $^{111}$Cd$^+$, the relevant energy splittings are: $\gamma / 2\pi=60 \textrm{~MHz}$, $\Delta_1/2\pi=13.7 \textrm{~GHz}$, $\Delta_2/2\pi=800 \textrm{~MHz}$ \cite{Tanaka96,Moehring06} and we choose the laser wavelength such that the detuning $\delta \approx 0$ (figure~\ref{dipolestrengths}). The branching ratios are calculated from eqns.~\ref{m1ratio}, \ref{m2pratio}, and \ref{m2sratio}: ${\bf M_1}=2/9$ and ${\bf M_{2\pi}} = {\bf M_{2-}} = 1/9$. Because ${\bf M_{2\pi}} = {\bf M_{2-}}$, the $\pi$- and $\sigma^{-}$-polarized components can be treated together and we can define an overall laser power impurity $P_{impure}\equiv P_{\pi}+P_{-}$.  Experimentally, we know the detection time, $\tau_D$, and we can leave $\eta$, $P_{impure}$, and $s$ as free parameters to be determined by the fit to experimental data (although in principle one could independently measure these parameters). Figure~\ref{detectfit} shows the good agreement between theory and experiment.

\subsection{Theoretical detection fidelity for $^{111}$Cd$^+$}
\noindent
\subsubsection{Detection with $P_{3/2}$ levels} \label{theoryfidelity}
\noindent
To determine the theoretical limit of detection fidelity for $^{111}$Cd$^+$ using the $P_{3/2}$ levels we choose the most ideal conditions: small laser detuning from resonance ($\delta \to 0$) and perfect detection beam polarization so the bright state histogram is a true Poissonian ($P_{impure}=0$ which means $\alpha_2=0$). Moreover, with perfect polarization we can lower the laser intensity ($s\to0$) to eliminate power-broadening while increasing $\tau_D$ to maintain a sufficiently bright ``bright'' state.

Using a discriminator at photon level $d$, detection fidelity of the bright state is the probability that the ion scatters {\it more} than $d$ photons while for the dark state it is the probability that it scatters $d$ or fewer photons. The overall qubit detection efficiency is the lower of these two numbers and it is maximized for the two fidelities being equal. Thus, the optimal qubit detection fidelity is:
\begin{equation} \label{exactfidelity}
F=\sum_{n=0}^d p_{\mathrm{dark}}(n) = 1-\sum_{n=0}^d p_{\mathrm{bright}}(n)
\end{equation}
with $\lambda_0 = s \tau_D \eta \frac{\gamma}{2}$ and $\alpha_1 = {\bf M_1} (\frac{\gamma}{2\Delta_1})^2$ given our assumptions mentioned above.

To calculate the detection fidelity, $F$, we find $\alpha_1=1.1\times 10^{-6}$ for $^{111}$Cd$^+$. Then we can choose the optimal $\lambda_0$ for discrimination by controlling the effective light level on the ion (adjusting the product $s\tau_D$). For our current value of $\eta \approx 0.001$ the optimum light level yields $\lambda_0=5.6$ and, using the discriminator level $d=0$, we obtain a qubit detection fidelity of $F=99.5\%$. In practice the optimal light level is slightly higher to aid in discriminating the background laser scatter from the bright qubit fluorescence.

\begin{figure} [htbp]
\centerline{\epsfig{file=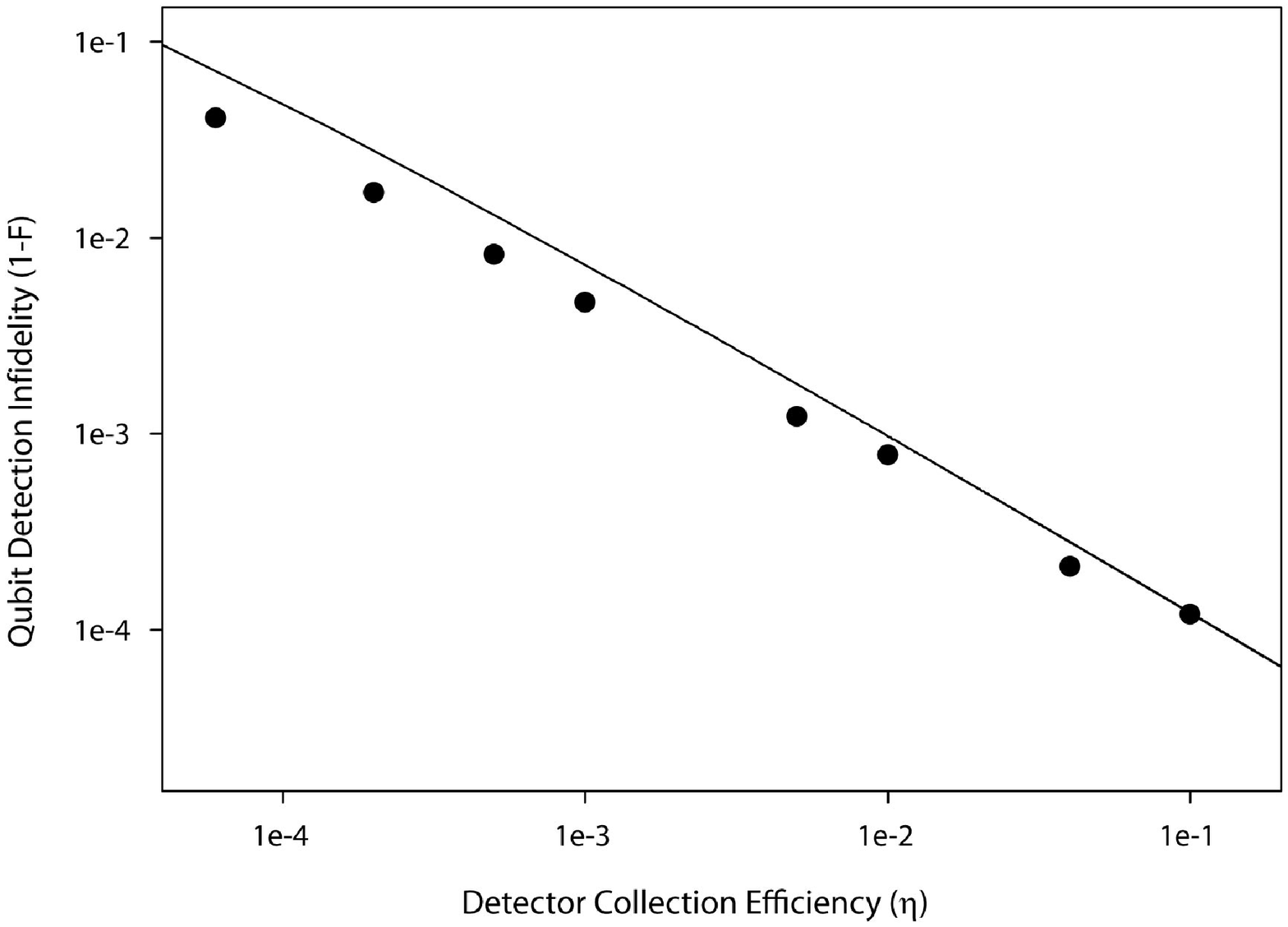, width=3.0in}}
\vspace*{13pt}
\fcaption{\label{detectpredict}Theoretical prediction of detection infidelity (1-F) using the $P_{3/2}$ detection scheme for $^{111}$Cd$^+$ as a function of total detector collection efficiency ($\eta$). Filled circles are numerically calculated using eqn.~\ref{exactfidelity} with iterations to find the optimal light level \& detection time for each $\eta$; solid curve uses the approximate analytic result of eqn.~\ref{approxfidelity}.}
\end{figure}

We can obtain a useful approximate analytic result for the theoretical qubit fidelity as a function of detector collection efficiency by assuming a discriminator level $d=0$. The optimal situation is when the fidelity of bright state detection and dark state detection are equal: $F = e^{-\alpha_1 \lambda_0/\eta}=1-e^{-\lambda_0}$ (eqn.~\ref{exactfidelity}). The term $\alpha_1 \lambda_0/\eta$ is typically small so we Taylor expand and then take the natural logarithm of both sides to find:
\begin{equation}
\lambda_0+\ln \lambda_0 \approx -\ln (\alpha_1/\eta)
\end{equation}
We wish to obtain a closed form solution for $\lambda_0$ as a function of $\alpha_1$ and $\eta$ since $\alpha_1$ depends only on atomic parameters. Therefore, for the typical case of optimal $\lambda_0 \sim 5-15$ it is reasonable to further assume that $\ln \lambda_0 \ll \lambda_0$ so that the ideal light level is approximately $\lambda_0 \approx -\ln (\alpha_1/\eta) = \ln (\eta/\alpha_1)$ and the approximate fidelity is:
\begin{equation} \label{approxfidelity}
F_{\mathrm{approx}}\approx1-\frac{\alpha_1}{\eta}\ln (\frac{\eta}{\alpha_1}).
\end{equation}
This result gives an accurate scaling of the fidelity as a function of detector collection efficiency (figure~\ref{detectpredict}).

In an ideal case, we could use a cavity to surround the ion \cite{Keller04} to increase the solid angle of light collected and a detector with very high quantum efficiency. This might produce a photon collection efficiency as high as $\sim30\%$ which, using eqn.~\ref{exactfidelity}, would yield a fidelity: $F_{\mathrm{ideal}}=99.997\%$.

At this point we hit a fundamental limitation of fluorescence state detection when using the $P_{3/2}$ levels and a qubit stored in the clock states: even with perfect polarization the $\ket{1,0}\equiv \ket{{\bf 1}}$ state can off-resonantly couple to the $\ket{0,0}\equiv \ket{{\bf 0}}$ dark state before ever reaching the cycling transition. This coupling is unaffected by any external parameters and is a property of the atomic structure only. Given this inherent error, the maximum fidelity of direct clock state qubit detection is given by:
\begin{equation} \label{cyclingerror}
F_{\mathrm{max}}=1-P_{\ket{1,0}\to \mathrm{dark}} = 1-\frac{4}{9}\left(\frac{\gamma}{2\omega_{\mathrm{HFP}}}\right)^2
\end{equation}
where the factor of $4/9$ comes from the squared Clebsch-Gordon coefficients for transitions out of the $\ket{{\bf 1}}$ state in Cd$^+$. For ions where the $P_{3/2}$ hyperfine splitting is not much larger than the linewidth, the ``bright'' clock state is not actually very bright because it couples to the dark manifold so easily. Direct fluorescence state detection of clock states is basically impractical for these ions, but there are a number of schemes to improve detection fidelity including using Raman transitions to shelve one of the qubit states in a more off-resonant state (with the detection fidelity now possibly limited by the Raman $\pi$-pulse fidelity) and using large magnetic fields to make 1st-order insensitive states with $m_F\neq0$ \cite{Langer05}. For $^{111}$Cd$^+$, $\omega_{\mathrm{HFP}}\gg \gamma/2$, so direct clock state detection yields $F_{\mathrm{max}}=99.94\%$. Reaching this limit requires a high quantum efficiency detector (see eqn.~\ref{approxfidelity}), a subject we address in sec.~\ref{future}.

\subsubsection{Detection with $P_{1/2}$ levels}
\noindent
The theoretical limit on detection fidelity when coupling to the $P_{1/2}$ levels is the leakage from one of the bright states into the dark state. With no laser detuning ($\delta \to 0$) and low light level ($s\to0$), the relevant parameters become: $\lambda_0'=s\tau_D \eta \frac{\gamma'}{2}$, $\alpha_1'=\frac{2}{9}\left(\frac{\gamma'}{2\Delta_1'}\right)^2$, and $\alpha_2'=\frac{2}{9}\left(\frac{\gamma'}{2\Delta_2'}\right)^2$. It is always true that $\Delta_1'>\Delta_2'$, so $\alpha_2'>\alpha_1'$ which means that there is always more leakage from the bright state into the dark state than vice versa. Table~\ref{detecttable} provides the relevant energy splittings for $^{111}$Cd$^+$, $^{171}$Yb$^+$, and $^{199}$Hg$^+$ and the calculated detection fidelity for given values of $\eta$, the detector collection efficiency\footnote{Note that the fidelity for detecting the qubit state in $^{171}$Yb$^+$ is slightly more complicated because there is an allowed decay from the excited $P_{1/2}$ state to a low-lying $D_{3/2}$ state which can be re-pumped to the $S_{1/2}$ level via the $[3/2]_{1/2}$ level \cite{Tamm00}. This repumping step is identical to the standard detection step, but its infidelity can be ignored beacuse the off-resonant coupling is much smaller than for the $P_{1/2}$ detection step due to the energy splittings: $\gamma_{[3/2]}/2\pi=9.5$~MHz, $\Delta_{\mathrm{HFD}}/2\pi=0.86$~GHz, $\Delta_{\mathrm{HF[3/2]}}/2\pi=2.5$~GHz, and $\lambda=935.2$~nm.}. Note that for $^{111}$Cd$^+$ the advantage of using the $P_{1/2}$ detection scheme instead of the $P_{3/2}$ scheme is only realized for large values of $\eta$.

\vspace*{4pt}   
\begin{table}[hb]
\tcaption{\label{detecttable}Relevant energy splitting parameters for the $I=1/2$, $P_{1/2}$ detection scheme using $^{111}$Cd$^+$ \cite{Tanaka96,Moehring06}, $^{171}$Yb$^+$ \cite{Engelke96}, and $^{199}$Hg$^+$ \cite{Raizen92} with detection fidelities calculated using eqn.~\ref{brightprobfinal} for varying detector collection efficiency $\eta$.}
\centerline{\footnotesize\smalllineskip
\begin{tabular}{c|c|c|c}
& $^{111}$Cd$^+$ & $^{171}$Yb$^+$ & $^{199}$Hg$^+$ \\
\hline
$\lambda (S_{1/2} \to P_{1/2})$ (nm) & 226.5 & 369.5 & 194 \\
$\gamma'/2\pi$ (MHz) & 50 & 23 & 70 \\
$\Delta_{\mathrm{HFS}}/2\pi$ (GHz) & 14.5 & 12.6 & 40.5 \\
$\Delta_{\mathrm{HFP}}'/2\pi$ (GHz) & 2 & 2.1 & 6.9 \\
$\eta=0.001$ & $F=96.7\%$ & $F=99.33\%$ & $F=99.43\%$\\
$\eta=0.01$ & $F=99.65\%$ & $F=99.93\%$ & $F=99.943\%$\\
$\eta=0.3$ & $F=99.988\%$ & $F=99.998\%$ & $F=99.998\%$ \\
\end{tabular}}
\end{table}

\section{Individual Ion Detection Using a CCD}
\noindent
\subsection{CCD technical overview}
\begin{figure} [htbp]
\centerline{\epsfig{file=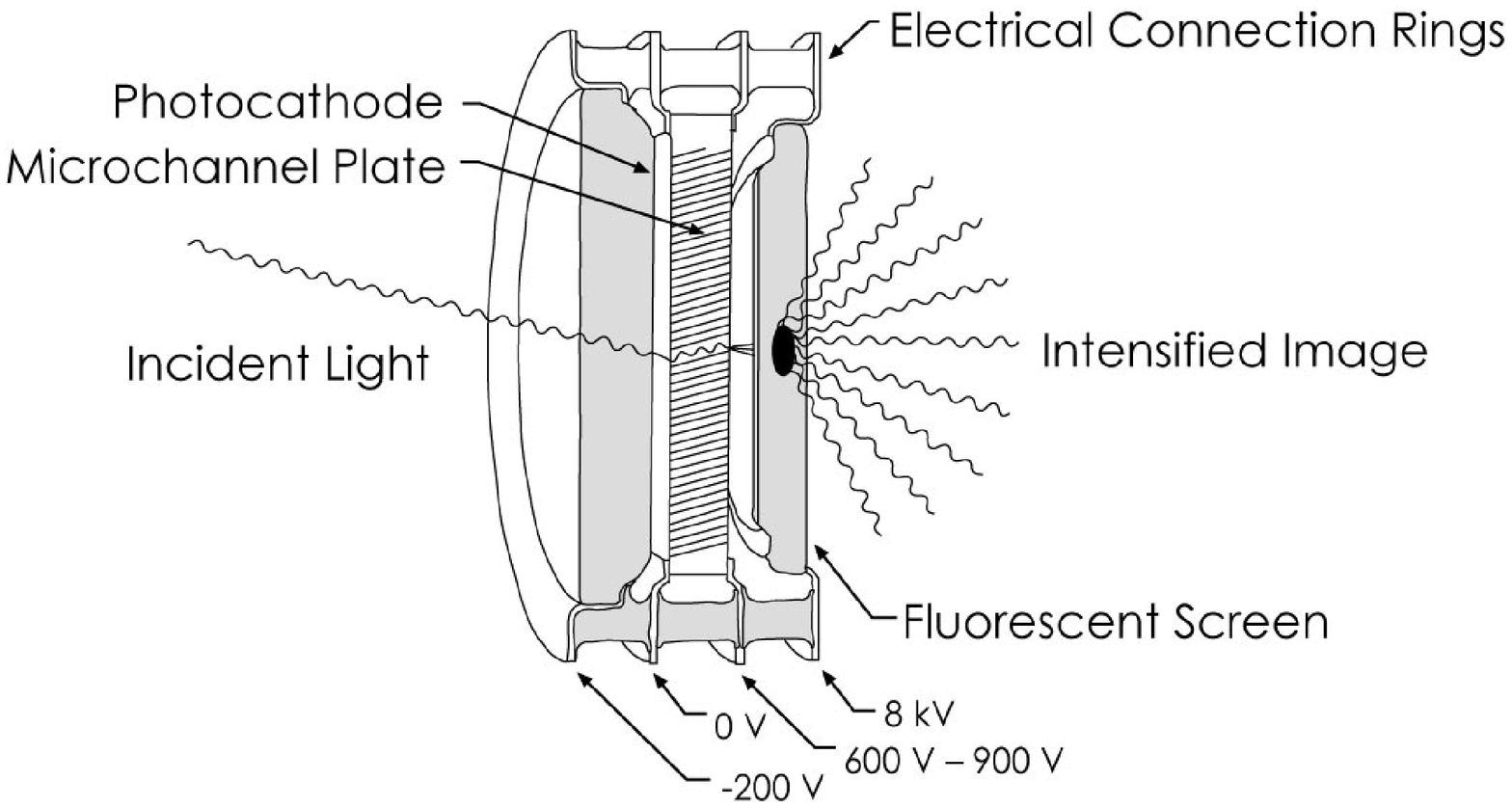, width=3.0in}}
\vspace*{13pt}
\fcaption{\label{CCDdiagram}Schematic diagram of an intensified CCD camera imaging tube. Incident photons impact the photocathode where they are converted to electrons and accelerated across the multichannel plate. These accelerated electrons strike the phosphor screen and produce visible photons that are coupled into fiber optics and fed onto a standard CCD for readout. Diagram used with permission of Princeton Instruments / Acton.}
\end{figure}

\noindent
To benefit from the spatial resolution of a CCD one must typically use an intensified CCD to obtain a signal that is much larger than the CCD readout noise (eqn.~\ref{SNR})\footnote{Recent advances in electron-multiplying CCD (EMCCD) architectures allow for single-photon detection at high readout speeds with low noise (although with no gating capability) \cite{emccd}. The following discussion, in particular eqn.~\ref{SNR}, is equally valid for EMCCDs by using the appropriate gain and quantum efficiency.}. In an intensified CCD, single photons incident on the front-screen photocathode produce electrons that are accelerated across a multichannel plate to start a localized electron avalanche that impacts a phosphor screen (figure~\ref{CCDdiagram}). Negative (positive) biasing of the photocathode relative to the multichannel plate produces significant (no) electron acceleration; this differential effect allows for rapid gating of the intensifier and reduces background counts. Visible wavelength photons emitted from the phosphor are then coupled (via lenses or fiber optics) to a standard CCD that converts the photons to charge for readout. The charge on each pixel of the CCD is then measured with the value proportional to the incident light intensity. A computer PCI board converts the analog voltage signal for each pixel to a digital integer for computer processing.

Technical noise considerations dictate the important CCD characteristic of readout speed. For a qubit in the bright state, the average number of electrons produced at the photocathode by photons incident on the detector during detection is: $\lambda_0=n_{ion} \eta_D T\textrm{d}\Omega/4\pi$ where $n_{ion}$ is the number of photons emitted by the ion, $\eta_D$ is the quantum efficiency of photon to electron conversion, $T$ is the optical transmission between the ion and the detector, and $\frac{\textrm{d}\Omega}{4\pi}$ is the solid angle of light collection. The remaining stages of electron intensification, conversion to photons, and then photon-induced charge production can be summarized by a gain factor $g$ so that the CCD rms shot noise will be: $g\sqrt{\lambda_0}$. To readout the charge on a pixel, the CCD controller must first clear the charge accumulated during the previous pixel readout. The imperfect repeatability of this process will induce some noise on the signal, with faster readout speeds leading to less perfect charge cleaning and therefore more noise. This rms noise per readout, $r$, is uncorrelated to the shot noise so the two noise sources add in quadrature to produce a total noise: $\sqrt{g^2 \lambda_0 + (kr)^2}$ where $k$ is the total number of pixels readout. Therefore, the total signal-to-noise ratio (SNR) becomes:
\begin{equation} \label{SNR}
\textrm{SNR}=\frac{g \lambda_0}{\sqrt{g^2 \lambda_0 + (kr)^2}}=\frac{\lambda_0}{\sqrt{\lambda_0 + (\frac{kr}{g})^2}}.
\end{equation}

Many CCDs allow on-chip binning of multiple pixel charges together before readout. This on-chip binning increases readout speed and decreases readout noise (because less pixels are readout) at the expense of decreasing spatial resolution (although in principle resolutions above one pixel per ion are unnecessary for distinguishing between the bright and dark states). If $\sqrt{\lambda_0}\ll kr/g$ then the signal is readout noise limited and $\textrm{SNR}\approx \frac{g \lambda_0}{kr}$, so the SNR will improve linearly with the number of pixels binned. For timescales typical of an ion fluorescence experiment, $kr\approx10$ and so using an intensifier with $g \gg 1$ allows one to use the gain to overwhelm the readout noise. In this shot noise limited regime, the signal-to-noise ratio becomes $\textrm{SNR}\approx \sqrt{\lambda_0}$.

\subsection{CCD experimental usage}
\noindent
For efficient experimental detection we use a $28\times28$ pixel box on-chip binned $4\times4$ such that there is an effective box size of $7\times7$ ``super''-pixels. This box size was chosen to match the imager-magnified ion-ion spacing in a trap with $\omega_z/2\pi=2.0\textrm{~MHz}$ so that each ion would be centered in its box with no overlap of boxes; the $4\times4$ binning improves readout speed yet still offers enough imaging resolution to provide real-time monitoring of ion/imaging system drifts. The detection signal is formed from the integrated electron counts of these 49 pixels minus the constant offset due to the non-zero readout charge maintained on each CCD pixel. Note that this offset is {\em not} the same as the negligible CCD dark counts that are caused by inadvertent electron transfer to the back-screen due to thermal effects. We are able to significantly suppress this thermal effect by cooling the CCD. When histogrammed, the integrated counts for $n$ incident photons shows a distribution because multiple electrons and photons are produced for each incident photon with a mean number of integrated counts for this box size of $\sim 100$ integrated counts per incident photon. Experimentally, the optimal discrimination level is chosen by equalizing the fraction of misidentified states, i.e. choosing a discrimination level such that an equal fraction of $\ket{1,0} \equiv \ket{{\bf 1}}$ is misidentified as ``dark'' and $\ket{0,0} \equiv \ket{{\bf 0}}$ is mislabeled as ``bright''. Using this optimal discriminator the efficiency of detecting a single ion's quantum state is $>99.4\%$ with the lower detection fidelity attributable to the CCD's smaller light collection angle (the PMT counts all the incident light while the CCD only uses counts inside the ion pixel box).

\section{Multiple Ion Detection Using a CCD}
\noindent
\subsection{Experimental implementation}
\noindent
Of more practical interest than single-ion detection is the ability to detect multiple ion quantum states simultaneously using the CCD's spatial resolution. For each ion we select a box or region of interest (ROI) that determines the pixels over which the CCD will integrate to determine the ion's state. Ion ROI box edges are positioned on adjacent pixels with no overlap to maximize detection fidelity and on-chip binning is used to enhance readout speed.

\begin{figure} [htbp]
\centerline{\epsfig{file=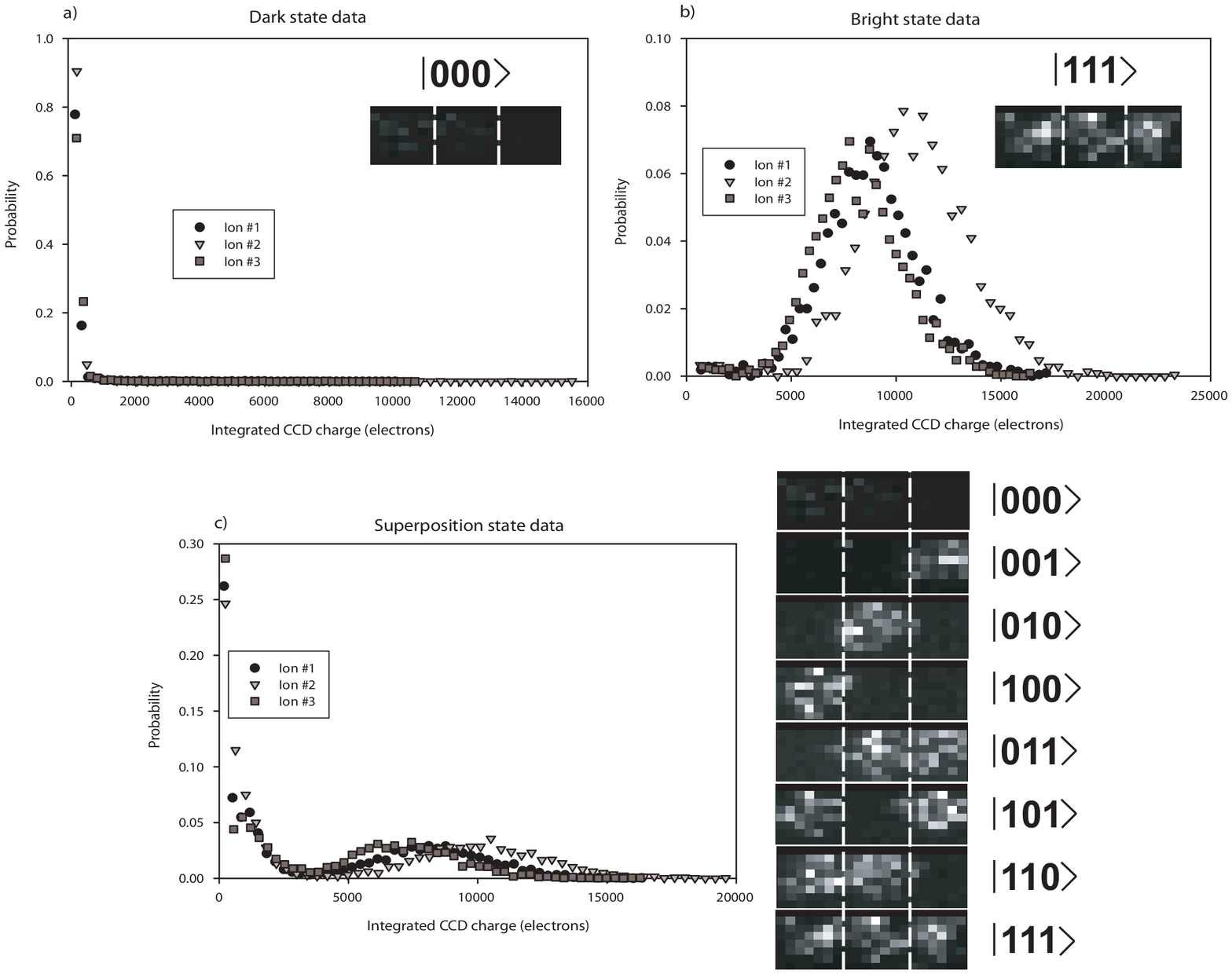, width=\columnwidth}}
\vspace*{13pt}
\fcaption{\label{3ionhistograms}Detection histograms for three ion qubits in the states: a) $\ket{0,0} \equiv \ket{{\bf 0}}$; b) $\ket{1,0} \equiv \ket{{\bf 1}}$; and c) an equal superposition state prepared by starting in $\ket{{\bf 000}}$ and applying a microwave $\pi/2$-pulse equally on all three qubits to produce the state $(\ket{{\bf 0}} + \ket{{\bf 1}})^{\otimes 3}$. Each graph contains 4000 trials. Note that ion \#2 has the most integrated counts due to unequal illumination by the detection beam. Adjacent to each histogram are examples of the post-selected single-shot images acquired by the CCD for each case: a) all ``dark'' ($\ket{{\bf 000}}$); b) all ``bright'' ($\ket{{\bf 111}}$); and c) all combinations of ``dark'' and ``bright''. The dashed white lines indicate the boundaries of the regions of interest used to determine the qubit state via integrated CCD counts.}
\end{figure}

Figure~\ref{3ionhistograms}a and b shows detection histograms and the actual CCD ``pictures'' for three ions prepared in the $\ket{{\bf 000}}$ dark state and the $\ket{{\bf 111}}$ bright state. In fig.~\ref{3ionhistograms}c we show the detection histograms and post-selected examples of all possible pictures for an equal superposition state achieved by preparing the state $\ket{{\bf 000}}$ and then applying a microwave $\pi/2$-pulse at the hyperfine splitting $\omega_{\mathrm{HFS}}/2\pi=14.5\textrm{~GHz}$ equally on the three ions to produce the state $(\ket{{\bf 0}} + \ket{{\bf 1}})^{\otimes 3}$.

\subsection{Possible additional errors}
\noindent
In addition to the single-ion detection errors due to off-resonant coupling (sec.~\ref{detecttheory}), simultaneous multiple ion detection on a CCD can produce some additional errors. One possible error is that the detection beam has a waist of $\sim10~\mu$m, compared to an ion-ion spacing of $\sim4~\mu$m for this experiment, leading to unequal illumination of the ions (visible in the histograms of fig.~\ref{3ionhistograms} where the middle ion is brighter than the outer two ions). In general, this unequal illumination results in different constant offsets for each ion's light level that are easy to correct.

The more significant error source is the potential for adjacent ion effects, namely optical pumping from ion-to-ion and box-to-box leakage on the CCD. Pumping an ion into the ``bright'' state from the fluorescence of the adjacent state is completely neglible compared to the effect of the laser. The light intensity due an adjacent bright ion is: $\frac{\hbar \omega \gamma/2}{4\pi x^2}$ where $\omega$ is the frequency of the radiated light and $x$ is the inter-ion spacing. In comparison, the laser intensity is typically: $I_{\mathrm{laser}}\approx I_{\mathrm{sat}}=\frac{\pi \hbar c \gamma}{3 \lambda^3}$ where $\lambda$ is the wavelength of the radiated light. Thus, the light intensity effect of the adjacent ion relative to the laser is:
\begin{equation}
\frac{I_{\mathrm{ion}}}{I_{\mathrm{laser}}}=\frac{3\lambda^2}{4\pi x^2}.
\end{equation}
For $^{111}$Cd$^+$, $\lambda=214.5$~nm and with three ions in a trap with $\omega_z/2\pi \approx 850$~kHz, $x\approx 4~\mu$m. Thus, $I_{\mathrm{ion}}/I_{\mathrm{laser}}\approx 7\times10^{-4}$, which is completely negligible.

Box-to-box leakage on the CCD can be most clearly seen by examining the conditional probabilities of the ions dependent on the state of the other ions. By independently rotating the qubits with microwave pulses, there should be no correlation between qubit states, but light leakage between boxes would produce such correlations. Using the data of figure~\ref{3ionhistograms} we find that adjacent ion pairs (left/center and center/right) exhibit conditional probability correlations of $\sim1.2\%$. With this leakage, the detection fidelity for individual qubits in the presence of other qubits is $\sim 98\%$. This fidelity is fully consistent with a small amount of leakage of the light from one ion into the ROI of the adjacent ion, an effect most likely caused by ion/imaging system drifts over a few minutes. Further mechanical stabilization of the imaging system should eliminate this leakage.

\section{Future Work} \label{future}
\noindent
The current detection fidelity of $\sim 98\%$ is highly efficient, but improvements in light collection and/or CCD quantum efficiency would be necessary to increase the fidelity to perform fault-tolerant quantum computing with a reasonable number of qubits. In addition, the CCD total readout time of $\sim15$~ms is much longer than a typical gate time of $\tau_{gate}\sim 100\mu$s, limiting the CCD's effectiveness for algorithms with steps conditional upon state detection (such as quantum error-correction \cite{Chiaverini04} or the quantum Fourier transform \cite{Chiaverini05}). Both limitations are ultimately technical. Current state-of-the-art unclassified CCDs operating in the near-IR can produce readout speeds of $\sim10\textrm{~MHz}$ per pixel or $\sim2.5~\mu$s per ion ROI with quantum efficiencies of $\sim60\%$. With these types of performance characteristics, employing CCDs for simultaneous multiple ion detection in feedback algorithms will be practical and efficient. Note that the fundamental limit on detection speed is given by the lifetime of the excited state: $\tau = \frac{2\pi}{\gamma}$. Thus, with high-efficiency light collection, the state detection time for Cd$^+$ could be as short as $\sim50-200$~ns.

Large-scale ion trap arrays have been proposed \cite{Kielpinski02,Slusher05} and small, scalable traps have been successfully produced using microfabrication techniques \cite{Stick06,Seidelin06}. Integrating multiple qubit detection via the CCD with these traps would produce a truly scalable qubit processing architecture.

\end{document}